\documentclass[amsmath,notitlepage,amssymb,aps,showkeys,floatfix,prd,a4paper,nofootinbib]{revtex4-1}

\usepackage{adjustbox}
\usepackage{amsmath}
\usepackage{graphicx}
\usepackage{epstopdf}
\usepackage{amsfonts,amssymb}
\usepackage{float}
\usepackage[utf8]{inputenc}
\usepackage{pstricks}
\usepackage{color}
\usepackage[compat=1.0.0]{tikz-feynman}
\usepackage{simplewick}
\usepackage{color, colortbl}
\definecolor{Gray}{gray}{0.9}
\usepackage{float}
\usepackage{tikz-feynman}
\usepackage{feynmp}
\usepackage{forest}

\usepackage{verbatim}    
\usepackage{dcolumn}
\usepackage{bm}
\usepackage{epsfig}
\usepackage{braket}
\usepackage[normalem]{ulem} 
\usepackage{longtable,booktabs}
\usepackage{booktabs}

\usepackage{multirow}
\usepackage{verbatim}
\usepackage{hyperref}
\usepackage{tabularx}
\usepackage{bm}
\usepackage{epsfig}
\usepackage{braket}
\usepackage{longtable,booktabs}
\usepackage{tikz}
\usetikzlibrary{tikzmark}
\usepackage[normalem]{ulem}
\usepackage{stackengine}
\newcommand{\be}{\begin{equation}}
\newcommand{\ee}{\end{equation}}
\newcommand{\ben}{\begin{eqnarray}}
\newcommand{\een}{\end{eqnarray}}

\begin{document}

\title{Fully-heavy tetraquarks in the vacuum and in a hot environment}

\author{Vin\'icius S. Silva}
\email{viniciussouza@ufba.br}
\affiliation{Instituto de F\'isica, Universidade Federal da Bahia, Campus Universit\'ario de Ondina, 40170-115, Bahia, Brazil}
\author{C\'assio Pigozzo}
\email{cpigozzo@ufba.br}
\affiliation{Instituto de F\'isica, Universidade Federal da Bahia, Campus Universit\'ario de Ondina, 40170-115, Bahia, Brazil}
\author{Luciano M. Abreu}
\email{luciano.abreu@ufba.br}
\affiliation{Instituto de F\'isica, Universidade Federal da Bahia, Campus Universit\'ario de Ondina, 40170-115, Bahia, Brazil}

\begin{abstract}

We study the thermal behavior of quarkonia and fully-heavy tetraquark states associated to the  charmonium, bottomonium and bottom-charmonium mass spectra. The starting point is the Schrödinger formalism with a vacuum Cornell-like potential. The spin-spin, spin-orbit and tensor contributions are also considered to describe the structure of the vacuum quarkonia $Q\bar Q$ spectra ($Q$ denoting $c,b$ quarks). The parameters of the model are fixed using the experimental data of the $Q\bar Q$ states.  After that, this formalism is extended to the fully-heavy tetraquark states within  the $1^3 S_1$ axial diquark--$1^3 S_1$ axial antidiquark configuration $[QQ] [\bar Q \bar Q]$, and their vacuum mass spectra are obtained and compared to the experimental data recently obtained. Our predictions support the interpretation of the
$X(6600)$ (or $X(6552)$), $X(6900)$ and $X(7200)$ states as the radially-excited $T_{4c}(n^1S_0)$ configurations with $n=2,3,4$.
In the sequence, we evaluate the mass spectra behavior in a thermal medium, by introducing a modified temperature-dependent Cornell potential. As a consequence, this formalism enables us to get some insight into the dissociation mechanism of $[QQ] [\bar Q \bar Q]$ states caused by a thermal medium, and into the temperature range at which the tetraquark states might be formed. We find that these structures cannot be formed in the thermal medium when the system has a temperature higher than about twice the critical temperature. These findings may be useful to better understand the features of the exotics in heavy-ion collisions. 

\end{abstract}



\maketitle

\section{Introduction}



Recently, analyses of proton-proton collision data collected by the LHCb, ATLAS and CMS Collaborations in di-charmonia channels found sharp peaks consistent with new resonances. First, the LHCb collaboration reported a narrow structure in the $J/\psi-J/\psi$ invariant mass spectrum~\cite{Aaij:2020fnh}. This new resonance, denoted as $X(6900)$, was later confirmed by the ATLAS~\cite{ATLAS:2023bft} and CMS~\cite{CMS:2023owd} Collaborations. Additionally, the ATLAS also observed the new structures $X(6400)$ and $X(6600)$ in the di-$J/\psi$ channel and the $X(7200)$ in the $J/\psi-\psi(2S)$ spectrum~\cite{ATLAS:2023bft}; and the CMS identified the $X(6600)$ (or $X(6552)$), $X(6900)$ and $X(7200)$ in the $J/\psi-J/\psi$ channel as well~\cite{CMS:2023owd}. These states appear as viable candidates of fully-charm tetraquark states, $T_{4Q}$'s (for reviews, we refer the reader to Refs.~\cite{Chen:2022asf,Brambilla:2019esw}). Nevertheless, their quantum numbers $J^{PC}$ have not yet a definitive assignment. 

From theoretical side, the notion of $T_{4Q}$ has been investigated since the 1970's~\cite{Iwasaki:1975pv,Chao:1980dv,Ader:1981db,Silvestre-Brac:1992kaa,Esposito:2013fma,Carvalho:2015nqf,Eichten:2017ffp,Debastiani:2017msn,Lundhammar:2020xvw,Bedolla:2019zwg,Giron:2020wpx,Chen:2020xwe}, mostly by using different versions of quark models within the diquark--antidiquark configuration $[QQ] [\bar Q \bar Q]$ [where $Q (\bar Q)$ denotes $c,b (\bar c , \bar b)$ (anti)quarks]. However, the recent data above mentioned definitively stimulated a number of investigations of the $T_{4Q}$'s with $J^{PC}=0^{++},1^{+-},2^{++}$ in distinct frameworks (see e.g.~\cite{Chao:2020dml,Lu:2020qmp,liu:2020eha,Lu:2020cns,Wang:2020ols,Becchi:2020uvq,
Becchi:2020mjz,Karliner:2020dta,Yang:2020rih,Weng:2020jao,Mutuk:2021hmi,Dong:2020nwy,Cao:2020gul,Zhang:2020xtb,Wang:2020tpt,Faustov:2020qfm,Wu:2022qwd,Wang:2022xja,Goncalves:2021ytq,Biloshytskyi:2022dmo,Mutuk:2022nkw,Feng:2023agq,Agaev:2023wua,Wang:2023jqs,Ortega:2023pmr,Wu:2024euj,Meng:2024yhu,Abreu:2023wwg,Zhu:2024swp,Belov:2024qyi,Song:2024ykq}).
As demonstrated by these mentioned works, their production mechanisms and possible interpretation of their internal structure remains as a theme of intense debate. 

To shed light on the the properties of $T_{4Q}$'s, heavy-ion collisions (HICs) appears as a promising scenario (see for example discussions in~\cite{Abreu:2021jwm,Abreu:2022lfy,Abreu:2022jmi,Abreu:2023aqy,Abreu:2023awj,Abreu:2024mxc,Abreu:2024asn}). Nucleus-nucleus collisions are characterized by the phase transition 
from nuclear matter to a locally thermalized state of deconfined quarks and   
gluons (the quark-gluon plasma-QGP). Heavy quark-antiquark pairs are mostly produced in the early stage via hard scattering reactions, since their time scale is shorter than the QGP thermalisation time~\cite{Andronic:2015wma}. Although most of them hadronize into open heavy-flavor states, an amount of the produced pairs will form quarkonia and also exotic states like the $T_{4Q}$'s. In this context, the heavy-flavor and exotic production is expected to be significantly suppressed due to the color screening mechanism that weakens the binding interaction involving heavy quarks and antiquarks. 
Thus, the in-medium production and dissociation of these states play an relevant role in the analysis of the hot strongly interacting medium and might provide estimations of the temperatures characterizing the nucleus-nucleus collisions. 
This has given rise to studies concerning the exact melting point, mass shift and thermal width at which heavy states are expected to dissociate at sufficiently high temperature above the quark-hadron transition temperature, but a consensus has not been reached yet (see e.g.~\cite{Asakawa:2003re,Rothkopf:2019ipj,Kim:2018yhk,Song:2020kka}). In particular, we remark the recent analysis of the thermal behavior of the exotic state $X(3872)$, under the assumption that it is a tetraquark, through the use of a potential approach at finite temperature in the Born-Oppenheimer approximation~\cite{Armesto:2024zad}. It has claimed a picture for the dissociation of the $X(3872)$  state in a medium,  and obtained an estimation of the temperature range for its formation.


Thus, inspired by the discussion above, our purpose here is to investigate the thermal properties of the $T_{4Q}$'s in the  charmonium, bottomonium and bottom-charmonium mass spectra. We start by presenting the Schrödinger formalism with the presence of a vacuum Cornell-like potential. Contributions carrying the spin contributions are also included to account for the structure of the vacuum quarkonia spectra. The parameters of the model are fixed using the experimental data of the $Q\bar Q$ states.  After that, this formalism is extended to the $T_{4Q}$'s within the diquark--antidiquark configuration $[QQ] [\bar Q \bar Q]$, and their vacuum mass spectra are determined. Our predictions are compared to the available data from LHCb, CMS and ATLAS collaborations. Next, the quarkonium and $T_{4Q}$ mass spectra behavior in a thermal medium is then evaluated, by introducing a modified temperature-dependent Cornell potential. The dissociation mechanism of these states in the medium is analyzed and the temperature range of their formation is estimated. 

This paper is organized as follows. In the next section we will briefly describe the Schrödinger formalism and the vacuum potential employed in this work. The framework is then extended to describe the system in thermal medium, by means of a temperature-dependent potential. Section III is devoted to present the results of the fitting procedure for the quarkonia and the $T_{4Q}$ spectra in the vacuum, and also our predictions for the thermal behavior of the states under consideration.  Finally, in section IV we present some closing remarks.

\section{The non-relativistic quark model}


We start by introducing the quark model to be used. Keeping in mind that the systems treated here are quarkonia ($Q\bar Q$) and $T_{4Q}$'s with a  diquark--antidiquark configuration ($[QQ] [\bar Q \bar Q]$), we assume that both cases can be reduced to a two-body problem. Noticing that the momentum of the constituents should be very small compared to their rest mass, then it is reasonable to use a non-relativistic framework with static potentials. In this sense, we employ the time-independent Schr\"odinger equation incorporating spherically-symmetric potentials, which has the radial part written as:
\begin{equation}
 \left[-\frac{d^2}{dr^2} + V_{eff}\right]\phi(r) = 2 m E\phi(r),
 \label{eq_schro}
\end{equation}
where $\phi (r)= r \psi (r)$, with $\psi (r)$ being the wavefunction; $m = m_1 m_{2} / (m_1 + m_{2}) $ is the reduced mass ($ m_1 $ and $m_{2}$ being the masses of the constituent particles); $E$ is the energy eigenvalue related to the mass $M$ of the bound system by means of the expression
\begin{equation}
    M = m_1 + m_{2} + E; 
 \label{eq_mass1}
\end{equation}
$V_{eff}$ is the effective potential, given by
\begin{equation}
    V_{eff} = \frac{l(l+1)}{r^2} + V^{(0)}(r) ,
\label{eq_effpot1}
\end{equation}
with $l$ labeling the orbital angular momentum, and $ V^{(0)}(r) $ representing the non-perturbative contribution. Below we discuss in more detail the potentials encoding the two-body interactions considered in this work. 

\subsection{Potentials in the vacuum}

Following the seminal work~\cite{Barnes:2005pb}, we adopt a zeroth-order potential including  the standard color Coulomb plus linear scalar form, and a Gaussian-smeared contact hyperfine term coming from one gluon exchange (OGE) interactions. They are explicitly 
\begin{equation}
    V^{(0)}(r) = \kappa \frac{\alpha_S}{r} + br -\frac{8\pi\kappa\alpha_S}{3(2m)^2}\Big(\frac{\sigma}{\sqrt{\pi}}\Big)^3 e^{-\sigma^2r^2} \mathbf{S}_1 \cdot \mathbf{S}_{2} , 
\label{eq_cornell}
\end{equation}
where $\kappa$ is the color factor (see below), $\alpha_S$ is the QCD fine structure constant  and $b$ is the string tension. The spin operators $\mathbf{S}_1,  \mathbf{S}_{2} $ are associated with  the constituent particles. 

The other spin-dependent contributions are treated perturbatively and will yield mass shifts at leading order. They are the spin-orbit $V_{LS}$ and tensor $V_T$ terms~\cite{Barnes:2005pb,Debastiani:2017msn,Li:2019tbn}, written as
\begin{eqnarray}
V_{LS}(r) & = &  - \frac{1}{2m^2}\Big(3\frac{\kappa\alpha_S}{r^3} + \frac{b}{r}\Big) \ \mathbf{L}\cdot \mathbf{S} , \nonumber \\
V_{T} (r) & = &  - \frac{1}{m^2}\Big[ 3\frac{\kappa\alpha_S}{r^3}\Big] \ \left[ (\mathbf{S}_1\cdot \mathbf{\hat{r}})(\mathbf{S}_{2}\cdot \mathbf{\hat{r}}) - \frac{1}{3}(\mathbf{S}_1\cdot \mathbf{S}_{2})\right] ; 
     \label{vspin2}
\end{eqnarray}
where $\mathbf{S}$ the total spin operator and $\mathbf{L}$ the angular momentum operator.

The expectation value of the spin-dependent operators can be calculated using the proper diagonal basis. In this sense, the hyperfine (spin-spin) operator is determined in terms of the spin quantum numbers ${s, s_1, s_2}$, i.e. $  \langle \mathbf{S}_1 \cdot\mathbf{S}_2 \rangle = [s(s+1) - s_1(s_1+1) - s_{2}(s_{2} + 1)]/2$. 
The spin-orbit operator can be directly determined in its $| j, l, s \rangle$ diagonal basis (the number $j$ is associated with the total angular momentum $\mathbf{J}$), with the matrix elements being $\langle \mathbf{L}\cdot \mathbf{S} \rangle = [ j(j + 1)-l(l + 1)-s(s+1)]/2$. In the case of the tensor operator in Eq.~(\ref{vspin2}), we make use of the calculations reported in~\cite{Barnes:2005pb,Debastiani:2017msn}: taking into account that  $\mathbf{S}_1 ,  \mathbf{S}_{2}$ operators are associated to spin-$1/2$ particles, then the tensor operator gives nonvanishing a contribution only for diagonal matrix elements between  $L > 0$ spin-triplet states ($| j, l \neq 0, s=1 \rangle$); it reads 
\begin{equation}
     \langle (\mathbf{S}_1\cdot \mathbf{\hat{r}})(\mathbf{S}_{2}\cdot \mathbf{\hat{r}}) - \frac{1}{3}(\mathbf{S}_1\cdot \mathbf{S}_{2})\rangle_{S=1, l\neq 0} = \begin{cases}
         -\frac{l}{6(2l+3)},  j = l + 1;\\
         + \frac{1}{6}, \ \ \ \ \ \ \  j = l;\\
         -\frac{(l+1)}{6(2l-1)},  j = l-1.
     \end{cases}
     \label{tensor_op}
\end{equation}
In the end, for $S$-wave states $(l=0)$ the spin-orbit  and tensor terms do not contribute, being only relevant the zeroth-order potential.

\subsection{The Color Factor $\kappa$ and the, quarkonium, diquark and tetraquark states}

The color factor $\kappa$ is determined by invoking the fact that hadrons must be color singlets under the $SU(3)$ color symmetry of QCD. In other words, the resulting color state of the constituent quarks in a hadron must be irreducibly decomposed and associated with a unidimensional representation. Accordingly, starting with the quarkonium state $\vert Q \bar Q \rangle $,  in the fundamental color representation it can be decomposed as $\mathbf{3} \otimes \bar{\mathbf{3}} = \mathbf{1} \oplus \mathbf{8}$. As a consequence, for the color singlet state  we have $\kappa \ \propto - \frac{N_c^2 - 1}{2 N_c}$, and $N_c=3$ gives $ \kappa = - 4/3 $. 

Extending this formalism to the (anti)diquark system, in a fundamental $\mathbf{3} $-color representation  a diquark state  $\vert q  q \rangle $  is decomposed as $\mathbf{3} \otimes {\mathbf{3}} = \mathbf{\bar{3}}\oplus\mathbf{6}$, while an antidiquark  $\vert \bar q  \bar q \rangle $ obeys the decomposition $\mathbf{\bar{3}} \otimes \mathbf{\bar{3}} = \mathbf{3}\oplus\mathbf{\bar{6}}$~\cite{Lundhammar:2020xvw,Debastiani:2017msn,Jaffe:2004ph}. The (anti)triplet state has a corresponding color factor of $ \kappa = - 2/3 $ and its Coulomb part of the potential acquires an attractive character; whereas the (anti)sextet state has $ \kappa = + 1/3 $ and its Coulomb potential becomes repulsive.
Thus, the color-attractive and dominant antitriplet diquark and triplet antidiquark masses can be calculated in a similar way as for the quark-antiquark system,  just changing the color factor $ \kappa \to \kappa /2 $. In addition, following other works~\cite{Lundhammar:2020xvw,Debastiani:2017msn}, for convenience we also perform the change in the the string tension  $ b \to b /2 $. 

Hence, the antitriplet diquark and triplet antidiquark are employed as the constituents of a tetraquark $\vert [Q Q]_{\mathbf{\bar{3}} }  [\bar{Q}\bar{Q} ]_{\mathbf{3}} \rangle$. A four-body system is in turn factorized into a two-body problem like the quarkonia, whose decomposition  $ {\mathbf{3}} \otimes \mathbf{\bar{3}}  = \mathbf{1}\oplus\mathbf{8} $ yields the color singlet bound state with $\kappa = -\frac{4}{3}$, and mass given by
\begin{equation}
    M[4Q] = m_{QQ} + m_{\Bar{Q}\Bar{Q}} + E_{[QQ][\Bar{Q}\Bar{Q}]},
\end{equation}
where in $E_{[QQ][\Bar{Q}\Bar{Q}]}$ is implicit the zeroth and fist-order contributions. 

Another point worthy of mention is that since the color antitriplet diquark is antisymmetric in the color wavefunction, and has positive parity in the ground state,  the spatial and flavor wavefunctions should be symmetric, giving an antisymmeric total wavefuction and engendering the 
$n^{2s+1}l_j = 1^3 S_1$ as good quantum numbers for the diquarks~\cite{Jaffe:2004ph}. As a consequence, we consider here only ground-state axial diquarks. The quantum numbers of the  $T_{4Q}$'s should obey the relations for the parity and  charge conjugation $ P_T = (-1)^{l_T}$ and $ C_T = (-1)^{l_T + s_T}$ respectively, where $s_T, l_T$ denote the spin and angular momentum (for more details see~\cite{Debastiani:2017msn,Jaffe:2004ph}).

\subsection{Fitting Procedure}

The parameters of the model $(m_1, m_2, \alpha_S , b ,\sigma )$ are constrained by fitting them to the data of the $Q\bar Q$ spectra. Explicitly,  the mass in Eq.~(\ref{eq_cornell}) is determined by finding values of the free parameters that minimize the $\chi^2 $ function, defined as
\begin{equation}
    \chi^2 \equiv \sum_{i=1}^{N} \frac{ \left[ M_{i}^{(exp)}- M_{i}^{(theo)} \right]^2}{\Delta_i^2},
\end{equation}
where $N$ is the number of experimental data used, $M_{i}^{(exp)}$ is the experimental mass, $M_{i}^{(theo)}$ is the corresponding predicted mass obtained by our model, and $\Delta_i$ is the corresponding uncertainty. 

\subsection{The temperature-dependent potential}

Our interest here is on the properties of the tetraquarks in a hot strongly interacting medium formed in nucleus-nucleus collisions. Since the order of magnitude of the collision energy is associated with the temperature $ T $ of the medium, the potential model presented above will be generalized to include $T$ as a variable, in order to predict the in-medium masses of $T_{4Q}$'s.

Thus, inspired by previous works~\cite{Karsch:1987pv,Satz:2006uh,Song:2011xi,Gauss}
we rewrite the Cornell potential (the first two terms in the right side of Eq.~(\ref{eq_cornell})) in the generalized temperature-dependent version 
\begin{equation}
    V_C (r,T)= -\kappa \alpha_S \frac{e^{-m_D(T) r}}{r} + \frac{b}{m_D(T)} \left[ 1 -e^{-m_D(T) r} \right],
    \label{temp_cornell}
\end{equation}
where $m_D(T)$ is the Debye mass, given by~\cite{Gauss}
\begin{eqnarray}
        m_D (T) & = &  Tg(\Lambda)\sqrt{\frac{N_c}{3}+\frac{N_f}{6}} \nonumber \\
        & & + \frac{N_cTg(\Lambda)^2}{4\pi} \log{ \Bigg(\frac{1}{g(\Lambda)}\sqrt{\frac{N_c}{3}+\frac{N_f}{6}}\Bigg) } \nonumber \\
        & & + \kappa_1 T g(\Lambda)^2 + \kappa_2 T g(\Lambda)^2 , 
    \label{debye}
\end{eqnarray}
where $\Lambda = 2\pi T$ is the renormalization scale; $ N_f$ is the flavor number 
$\sigma_1 , \sigma_2$ are nonperturbative constants, 
fixed by a fit based on continuum corrected lattice results, yielding $\kappa_1 = 0.686 \pm 0.221, \kappa_2= -0.317 \pm 0.052$; and $g(\Lambda)$ the running coupling, for which it is used the result given in~\cite{Vermaseren:1997fq}, with the scale $\Lambda_{QCD} = 0.2145$ GeV.

In this scenario, it is also useful to introduce the so-called dissociation temperature of a given state, $T_{dis}$, which is the temperature above which the bound state no longer exists. To determine $T_{dis}$, we define the dissociation energy 
$\equiv E_b (T)$ as~\cite{Karsch:1987pv,Satz:2006uh,Song:2011xi,Gauss}
\begin{eqnarray}
E_b (T) & \equiv &  m_1 + m_2 + V_C (r \to \infty , T) - M(T) \\
& = & V_C (r \to \infty , T) - E(T) , 
\label{Eb}
\end{eqnarray}
where $E(T)$ is the $T$-dependent energy eigenvalue obtained from Eq.(\ref{eq_schro}) but with the Cornell potential replaced by $V_C (r,T)$. Accordingly, $T_{dis}$ is the value so that $E_b (T_{dis}) = 0$.
The temperature enters  through the $T$-dependence of the
Debye mass present in the $T$-dependent potential in Eq.~(\ref{temp_cornell}).

\section{Results}

\subsection{Potentials in the vacuum}

\subsubsection{Quarkonium states}

After introducing the model, in this section the obtained results will be presented and discussed. First, the parameters $(m_1, m_2, \alpha_S , b ,\sigma )$ are fitted to the experimental data from the PDG~\cite{pdg} as follows. We use three data sets: the first one (labeled as $C$) related exclusively to the charmonium sector of the spectrum (with 15 mesons); the second one  (labeled as $B$)  related exclusively to the bottomonium sector (with 10 mesons); and the third one (labeled as $BC$) related exclusively to the bottom-charmonium sector (with 3 mesons). In the fitting procedure, the parameters have uniform priors in ranges compatible to those chosen in Refs~\cite{Lundhammar:2020xvw,Debastiani:2017msn}: $0.05 \leq \alpha_S \leq 0.70, \ 0.01 \  \mathrm{GeV}^2 \leq b \leq 0.40 \ \mathrm{GeV}^2, \ 0.05 \  \mathrm{GeV} \leq \sigma \leq 1.6 \ \mathrm{GeV} , \ 1 \  \mathrm{GeV} \leq m_c \leq 1.9 \  \mathrm{GeV}, \  4 \  \mathrm{GeV} \leq m_b \leq 5 \  \mathrm{GeV} $, where $m_c, m_b $ denote the constituent masses of the charm and bottom quarks. Since our purpose relies on a qualitative analysis of the thermal behavior of the mesons and $T_{4Q}$'s, for the sake of computational economy we adopt an uncertainty $\Delta_i$ equivalent to a relative error of $10\%$.  

Tables~\ref{table1} and \ref{table1b} show the  experimental data sets used here and our predictions to their vacuum masses. The values of the parameters obtained from the minimization procedure for the different sectors are displayed in Table~\ref{table2}. In addition, the values of the reduced $\Tilde{\chi} ^2$ function are also shown ($\Tilde{\chi}^2 = \frac{\chi^2}{\nu} ; 
$ $\nu = N - k$ being the number of degrees of freedom, with $N$ being the number of data points and $k$ the number of free parameters). In the case of the bottom-charmonium sector, with only two established particles, we have used the same parameter set of the charmonium sector. Looking at the charmonium spectrum, which is the most studied sector, the values of the parameters as well as the results for the masses of the states are similar to those from Refs.~\cite{Lundhammar:2020xvw,Barnes:2005pb,Debastiani:2017msn,Li:2019tbn}. They are also in reasonable agreement with the experimental data, when we consider the uncertainties (not shown for the sake of conciseness) coming from different choices of the parameters, as displayed for $\alpha_s $ in Table~\ref{table2}, stressing that we focus on the qualitative analysis of the properties of the $T_{4Q}$'s.

\begin{table}[htbp!]
    \centering
    \caption{Experimental masses ($M^{(exp)}$) from the PDG~\cite{pdg} and the corresponding predicted vacuum mass ($M^{(theo)}$) obtained by our model for the mesons in the charmonium sector of the spectrum.}
    \label{table1}
\begin{tabular}{lccccc}
\toprule
\hline
$(Q\bar Q)$ & $M^{(0)}$ [GeV] & $\langle V_{LS}^{(1)} \rangle$ [GeV] & $\langle V_{T}^{(1)} \rangle$ [GeV] & $M_f^{2}$ [GeV] & $M^{\text{(Exp)}}$ [GeV] \\
\midrule
\hline
$\eta_c(1S)$   & 2.9924  & 0       & 0        & 2.9924  & 2.9839 \\
$J/\psi(1S)$   & $3.0917 $  & 0       & 0        & 3.0917  & 3.0969 \\
$\chi_{c0}(1P)$ & 3.5191   & -0.0639 & -0.0294  & 3.4258   & 3.4147 \\
$\chi_{c1}(1P)$ & 3.5191  & -0.0320 & 0.0147   & 3.5018  & 3.5107 \\
$h_c(1P)$      & 3.5105  & 0       & 0        & 3.5105   & 3.5254 \\
$\chi_{c2}(1P)$ & 3.5191  & 0.0320  & -0.0029  & 3.5481  & 3.5562 \\
$\eta_c(2S)$   & 3.6317  & 0       & 0        & 3.6317   & 3.6375 \\
$\psi(2S)$     & 3.6714  & 0       & 0        & 3.6714  & 3.6681 \\
$\psi(3770)$   & 3.7958  & -0.0088 & -0.0039  & 3.7831  & 3.7773 \\
$\psi_2(3823)$ & 3.7958  & -0.0029 & 0.0039   & 3.7967   & 3.8222 \\
$\psi_3(3842)$ & 3.7951  & 0       & 0        & 3.7951   & 3.8422 \\
$\chi_{c2}(3930)$ & 3.7958  & 0.0059  & -0.0011  & 3.8006   & 3.9222 \\
$\eta_c(3S)$ & 4.04808 & 0 & 0 &  4.04808 & - \\
$\psi(3S)$ & 4.07548 & 0 & 0 & 4.07548 & - \\
$\chi_{c0}(3P)$ & 4.29369 & -0.058405 & -0.0245876 & 4.2107 & - \\
$\chi_{c1}(3P)$ & 4.29369 & -0.0292025& 0.0122938& 4.27679 & - \\
$h_c(3P)$ & 4.28453 & 0 & 0 & 4.28453 & -  \\
$\chi_{c2}(3P)$ & 4.29369 & 0.0292025 & -0.00245876 & 4.32044 & - \\
$\eta_c(4S)$ & 4.39326 & 0 & 0 & 4.39326 & - \\
$\psi(4S)$ & 4.41494 & 0 & 0 & 4.41494 & - \\
$\chi_{c0}(4P)$ & 4.60446 &-0.0576786& -0.0237132 & 4.52307 & - \\
$\chi_{c1}(4P)$ & 4.60446 & -0.0288393 & 0.0118566 & 4.58748 & - \\
$h_c(4P)$ & 4.59557 & 0 & 0 & 4.59557 & - \\
$\chi_{c2}(4P)$ & 4.60446 & 0.0288393 & -0.00237132 & 4.63093 & - \\
\hline
\bottomrule
\end{tabular}
\end{table}


\begin{table}[htbp!]
    \centering
    \caption{Experimental masses ($M^{(exp)}$) from the PDG~\cite{pdg} and the corresponding predicted vacuum mass ($M^{(theo)}$) obtained by our model for the mesons in the bottomonium and bottom-charmonium sectors of the spectrum.}
    \label{table1b}
\begin{tabular}{lccccc}
\toprule
\hline
$(Q\bar Q)$ & $M^{(0)}$ [GeV] & $\langle V_{LS}^{(1)} \rangle$ [GeV] & $\langle V_{T}^{(1)} \rangle$ [GeV] & $M_f^{2}$ [GeV] & $M^{\text{(Exp)}}$ [GeV] \\
\midrule
\hline
$\eta_b(1S)$ & 9.4191 & 0 & 0 & 9.4191 & 9.3987\\
$\Upsilon(1S)$ & 9.45147 & 0&0 & 9.4515 &9.4603 \\
$\chi_{b0}(1P)$   & 9.9321  & -0.0254 & -0.0117  & 9.8949  & 9.93208 \\
$\chi_{b1}(1P)$   & 9.9321  & -0.0127 & 0.00589  & 9.9253  & 9.93208 \\
$h_{b}(1P)$       & 9.9283  & 0       & 0        & 9.9282  & 9.92827 \\
$\chi_{b2}(1P)$   & 9.9321  & 0.01272 & -0.00118 & 9.9457  & 9.93208 \\
$\Upsilon(2S)$ & 10.0435 & 0&0 & 10.0435 &10.0232 \\
$\chi_{b0}(2P)$  & 10.3128 & -0.02161 & -0.0098947 & 10.2813 & 10.3128 \\
$\chi_{b1}(2P)$  & 10.3128 & -0.0108 & 0.00494739 & 10.3070 & 10.3128 \\
$\chi_{b2}(2P)$  & 10.3128 & 0.0108  & -0.0009894 & 10.3226 & 10.3128 \\
$B_c(1S)$ & 6.1372 & 0 & 0 & 6.1372 & 6.2744 \\
$B_c(2S)$ & 6.7634 & 0 & 0 &6.7634 & 6.8712 \\
\hline
\bottomrule
\end{tabular}
\end{table}


\begin{table}[htbp!]
    \caption{Best-fit parameter values for $(m_1, m_2, \alpha_S , b ,\sigma )$ from each fit of data sets $C,B , BC$ to the model, and the corresponding value of the reduced $\Tilde{\chi}^2$. $C$ is related exclusively to the charmonium sector of the spectrum (with 15 mesons); $B$ is related exclusively to the bottomonium sector (with 10 mesons); $BC$ is related exclusively to the bottom-charmonium sector. In the last case, with only two established particles, we have used the same parameter set of the charmonium sector.}
    \label{table2}
    \centering
    \begin{tabular}{c c c c c c c}
    \hline
   Data set & $m_1$ [GeV] & $m_2$ [GeV] &   $\alpha_S$ &   $b$ [Ge$V^2$] & $\sigma$ [GeV] & $\Tilde{\chi}^2$ \\
       \hline
C &        1.4622  &  1.4622   & 0.5202
& 0.1463& 1.0831 & 0.0063   \\
B &   4.7596 & 4.7596 &  0.3740 & 0.2052 & 1.5439  & 0.0051 \\
BC &       4.7596  &   1.4622     & 0.5202 & 0.1463 &  1.0831 & 0.0047\\
        \hline
    \end{tabular}
\end{table}


\subsubsection{Tetraquark states}

Once the best-fit values for the free parameters of the model have been determined, we can calculate the diquark masses following the prescription detailed in previous section, taking into account the different color structure of the diquark system encoded by the replacement of the parameters $ \kappa \to \kappa /2 $ and $ b \to b /2 $. The parameter sets shown in Table~\ref{table2} are then employed to calculate the masses of the $ cc , bb $ and $bc$ diquarks. In this exploratory study, only ground-state axial vector ($n^{2s+1}l_j = 1^3 S_1$) diquarks are considered. The values for the diquark masses obtained are presented in Table \ref{table3}. The change of the color factor in the potentials engenders ground-state axial vector diquarks with higher masses when compared to those of the $S$-wave ground-state mesons. In particular, here we focus on the the case of the $cc(1^3S_1)$, since there are available experimental data for the di-charmonium sector which we can take as a basis of comparison. We see that the value of $m_{cc}$ found is consonance with that found in other works, as~\cite{Lundhammar:2020xvw,Debastiani:2017msn}. However, as pointed out in the Table VII of~\cite{Lundhammar:2020xvw}, the value of $m_{cc}$ presents sizable deviation in literature; depending on the framework used this discrepancy can be by about 400 MeV. Relativistic effects seem to yield larger diquark masses, while models based on QCD sum rules techniques generate smaller $m_{cc}$. As a consequence, since the existing estimates for $m_{cc}$ do not converge and we still do not have a definite calculation for it, we  take its value within the range considered in  Table \ref{table3}.

\begin{table}[htbp!]
\centering
 \caption{Results obtained for the masses of ground-state axial vector diquarks, considering the parameter sets in Table \ref{table2} obtained from the minimization procedure for the quarkonium spectrum shown in Table~\ref{table1}.}
\begin{tabular}{ c c c }
\hline
       Diquark  & $m_{QQ}$ [GeV]  \\
 \hline      
        $cc(1^3S_1)$  &  $3.1334 \pm 0.062668$ \\
        $bb(1^3S_1)$ & $9.6285 $ \\
        $bc(1^3S_1)$  & $6.3498$ \\
        \hline
    \end{tabular}
 \label{table3}
\end{table}

After characterizing the diquark states, the diquark and antidiquark pairs are then combined,  according to the discussion in previous section, in order to generate the mass spectra for the tetraquark states; they are presented in Table~\ref{tab:tetraquark_predictions}. The contributions coming from the spin-orbit and tensor potentials are small but sufficient to break the degeneracy of the states with $l=1$.

\begin{table}[htbp!]
\centering
\caption{Summary of theoretical predictions for tetraquark masses, including spin-dependent contributions \(\langle V_{LS} \rangle\) and \(\langle V_T \rangle\).}
\label{tab:tetraquark_predictions}
\begin{tabular}{ccccccc}
\toprule
\hline
$T_{4Q}$ &  $n^{2s+1}l_j$ & $M^{(0)}$ [GeV] & $\langle V_{LS} \rangle$ [GeV] & $\langle V_T \rangle$ [GeV] & $M$ [GeV] \\
\hline
\midrule
$T_{4c}$ &  $1^1S_0$    & 6.03337   &0       & 0     & 6.03337   \\
$T_{4c}$ &  $1^3S_1$ & 6.08137 & 0& 0 &6.08137\\
$T_{4c}$ &  $1^3P_0$  & 6.5935   & -0.05373 & -0.020483 & 6.51929   \\
$T_{4c}$ &  $1^3P_1$  & 6.5935   & -0.026867 & 0.0102417 & 6.57687   \\
$T_{4c}$   &  $1^1P_1$ & 6.58654  & 0        & 0         & 6.58654  \\
$T_{4c}$ &  $1^3P_2$ & 6.5935  & 0.02686  & -0.00204835 & 6.61831 \\
\midrule
$T_{4b}$  &  $1^1S_0$   & 18.7679  & 0       & 0     & 18.7679  \\
$T_{4b}$  &  $1^3S_1$  & 18.7834 & 0& 0 &18.7834 \\
$T_{4b}$ &  $1^3P_0$  &19.422  & -0.026864 & -0.0096396 & 19.3855 \\
$T_{4b}$ &  $1^3P_1$   & 19.422  & -0.0260834 & 0.0048198 & 19.4007 \\
$T_{4b}$   &  $1^1P_1$ & 19.4189   & 0        & 0         & 19.4189  \\
$T_{4b}$  &  $1^3P_2$   & 19.422  & 0.01321  & -0.00094728 & 19.4343 \\
\midrule
$T_{2b2c}$  &  $1^1S_0$  & 12.0163  & 0       & 0        & 12.0163  \\
$T_{2b2c}$  &  $1^3S_0$ & 12.0371  &0 & 0& 12.0371 \\
$T_{2b2c}$ & $1^3P_0$   & 12.7724  & -0.21432& -0.075282 & 12.4828 \\
$T_{2b2c}$ & $1^3P_1$   & 12.7724  & -0.107165& 0.037641 &12.7029 \\
$T_{2b2c}$ & $1^1P_1$   & 12.7672  & 0& 0 & 12.7672  \\
$T_{2b2c}$ & $1^3P_2$   & 12.7724  & 0.107165& -0.0075282 & 12.6577 \\
\midrule
$T_{4c}$ &  $2^1S_0$    & 6.55988  &0& 0    & 6.55988   \\
$T_{4c}$ &  $2^3S_1$ &6.68904 & 0& 0 &6.68904\\
$T_{4c}$ &  $2^3P_0$  & 6.95699   & -0.044195  & -0.016568 & 6.8962   \\
$T_{4c}$ &  $2^3P_1$  & 6.95699   & -0.022097 & 0.008284 & 6.94318   \\
$T_{4c}$   &  $2^1P_1$ & 6.95144  & 0        & 0         & 6.95144  \\
$T_{4c}$ &  $2^3P_2$ & 6.95699  & 0.022097  & -0.0016568 & 6.97743 \\
\midrule
$T_{4b}$  &  $2^1S_0$   & 19.4883  &0       &0       &19.4883 \\
$T_{4b}$  &  $2^3S_1$  & 19.4911 & 0& 0 &19.491 \\
$T_{4b}$ &  $2^3P_0$  &19.7697  & -0.27014 & -0.0932131 & 19.4063 \\
$T_{4b}$ &  $2^3P_1$   & 19.7697  & -0.13507 & 0.0466066 & 19.6812 \\
$T_{4b}$   &  $2^1P_1$ & 19.7678   & 0        & 0         & 19.7678  \\
$T_{4b}$  &  $2^3P_2$   & 19.7697 & 0.13507  & -0.00932131 & 19.6253 \\
\midrule
$T_{2b2c}$  &  $2^1S_0$  & 12.8254  & 0        & 0       & 12.8254  \\
$T_{2b2c}$  &  $2^1S_0$ & 12.8291  & 0 & 0 & 12.8291 \\
$T_{2b2c}$ & $2^3P_0$  & 13.1165  & -0.157955& -0.0553793 & 12.9032 \\
$T_{2b2c}$ & $2^3P_1$   & 13.1165  & -0.0789775& 0.0276897 &13.0652 \\
$T_{2b2c}$ & $2^1P_1$   & 13.1138  & 0& 0 & 13.1138  \\
$T_{2b2c}$ & $2^3P_2$   & 13.1165  & 0.0789775& -0.00553793 & 13.032 \\
\midrule
$T_{4c}$ &  $3^1S_0$ & 6.91895 & 0 & 0 & 6.91895  \\
$T_{4c}$ &  $3^3S_1$ & 7.04278 & 0 & 0 & 7.04278  \\
$T_{4c}$ &  $3^3P_0$ & 7.24623 & -0.185573 & -0.0687683 &6.99189  \\
$T_{4c}$ &  $3^3P_1$ & 7.24623 & -0.0927876 & 0.0343845 & 7.18783  \\
$T_{4c}$ &  $3^1P_1$ & 7.24155 & 0 & 0 & 7.24155  \\
$T_{4c}$ &  $3^3P_2$ & 7.24623 & 0.0927876& -0.0343845 & 7.33214  \\
$T_{4c}$ &  $4^1S_0$ & 7.20517 & 0 & 0 & 7.20517  \\
$T_{4c}$ &  $4^1S_3$ & 7.32602 & 0 & 0 & 7.32602  \\
$T_{4c}$ &  $4^3P_0$ & 7.49774 & -0.175959 & -0.0646457 & 7.25713  \\
$T_{4c}$ &  $4^3P_1$ & 7.49774 & -0.0879654 & 0.0323179 & 7.44209  \\
$T_{4c}$ &  $4^1P_1$ & 7.49365 & 0 & 0 & 7.49365  \\
$T_{4c}$ &  $4^3P_2$ & 7.49774 & 0.0879795 & -0.00646357 & 7.57924  \\
\hline
\bottomrule
\end{tabular}
\end{table}


We can compare our predictions with those found in the literature. In general, our results are in convergence with those reported in the literature published before the discovery of the tetraquarks mentioned in the Introduction (see for example ~\cite{Lundhammar:2020xvw,Debastiani:2017msn}). 

\begin{figure}[htbp!]
    \centering
    \includegraphics[width=0.6\textwidth]{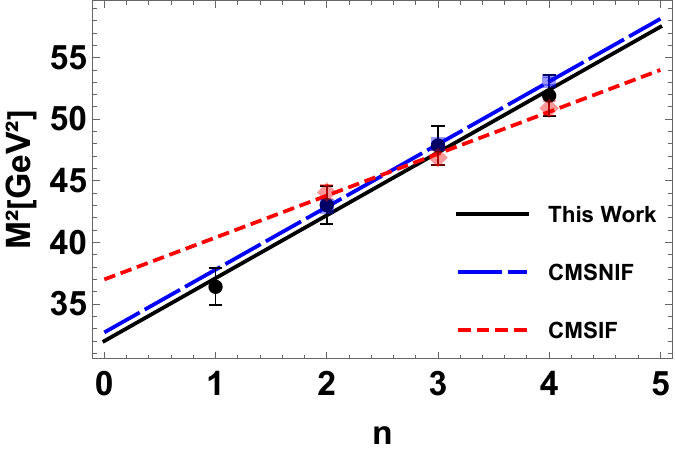}
    \caption{Regge trajectory in the $(n,M^2)$ plane for the $T_{4c}(n^1S_0)$ states  given in Tables~\ref{tab:tetraquark_predictions} and ~\ref{tab:tetraquark_predictions_new}, denoted as "Our". In addition, the Regge-like plot of the three resonances observed by the CMS Collaboration~\cite{CMS:2023owd} in Table~\ref{tab:tetraquark_predictions_new} are also shown. The lines are $\chi^2$ fits of the data, namely: $M_{(our)}^2(n)= 32.0+5.1 n, M_{(CMSNIF)}^2(n)= 32.7+5.1 n$, and $M_{(CMSIF)}^2(n)= 37.0+3.4 n$. }   
    \label{fig:regge}
\end{figure}

Most importantly, we now confront our findings with experimental data mentioned in the Introduction. Concentrating on the di-charmonium spectrum, one can notice that the radially-excited $T_{4c}(n^1S_0)$ states with $n=2,3,4$ are consistent with the structures $X(6600)$ (or $X(6552)$), $X(6900)$ and $X(7200)$ reported in the $J/\psi-J/\psi$ invariant mass spectrum between 6.2 and 9.0 GeV$/c^2$, yielded in proton-proton collisions at $\sqrt{13}$ TeV from data collected by the CMS Collaboration~\cite{CMS:2023owd}. To make the comparison easier, our predictions and the fitted Breit-Wigner masses and widths of these three resonances observed by the CMS are shown in Table~\ref{tab:tetraquark_predictions_new}. Accordingly, our findings suggest that these three structures are compatible with the fully-heavy tetraquark interpretation, with quantum numbers $0^{++}$, in particular with the case of no-interference fit. Interestingly, these results are in agreement with the speculation discussed in Ref.~\cite{Zhu:2024swp}, described as follows. From a Regge-trajectory analysis of the CMS data, the $X(6900)$ can be interpreted as a $n=3$ radially excited state, with the triplet masses being well-described by a typical linear relationship. Within this picture, the fundamental state with $n=1$ is still missing to be detected, but it would rely below the $J/\psi-J/\psi$ threshold in the no-interference fit, as in our case. This is better illustrated in Fig.~\ref{fig:regge}, where we see the Regge trajectory in $(n,M^2)$ plane for the $T_{4c}(n^1S_0)$ states given in Tables~\ref{tab:tetraquark_predictions} and ~\ref{tab:tetraquark_predictions_new}, together with the Regge-like plot of the three resonances observed by the CMS experiment. Hence, in this scenario the ground structure is below the threshold and still needs confirmation. Notwithstanding, it should be remarked that this predicted new state is also proposed in Refs.~\cite{Dong:2020nwy,Song:2024ykq}.


\begin{table}[htbp!]
\centering
\caption{Theoretical predictions for radially-excited $T_{4c}(n^1S_0)$ states ($n=2,3,4$) given in Table~\ref{tab:tetraquark_predictions}, and the fitted Breit-Wigner masses of the three resonances observed by the CMS Collaboration~\cite{CMS:2023owd}. The indices $NIF (IF)$ in the last two columns denote the data from no-interference (interference) fit. } 
\label{tab:tetraquark_predictions_new}
\begin{tabular}{ccccc}
\toprule
\hline
$T_{4Q}$ &  $n^{2s+1}l_j$ & $M$ [GeV] & $M^{\text{(Exp)}}_{NIF}$ [GeV]~\cite{CMS:2023owd} & $M^{\text{(Exp)}}_{IF}$ [GeV]~\cite{CMS:2023owd}  \\
\hline
\midrule
$T_{4c}$ &  $2^1S_0$   &   6.55988 $\pm 0.117449 $  & $6.552 \pm 0.010 \pm 0.012$  &  $6.638\substack{+0.043+0.016 \\ -0.038-0.031} $   \\
$T_{4c}$ &  $3^1S_0$   &  6.91895$\pm 0.116399 $  & $6.927 \pm 0.009 \pm 0.004$  & $6.847\substack{+0.044+0.048 \\ -0.028-0.020} $ \\
$T_{4c}$ &  $4^1S_0$   &  7.20517 $ \pm     0.11512$  & $7.287\substack{+0.020 \\ -0.018} \pm  0.005$ &  $7.134\substack{+0.048+0.041 \\ -0.025-0.015} $ \\
\hline
\bottomrule
\end{tabular}
\end{table}

\subsection{The temperature-dependent potential}

\subsubsection{Quarkonium states}

After obtaining the best-fit parameters for the system in vacuum, we now use the framework described earlier for the temperature-dependent potential to study how the hadronic states of our interest behave in the thermal medium. 

\begin{figure}[htbp!]
    \centering
    \includegraphics[width=0.6\textwidth]{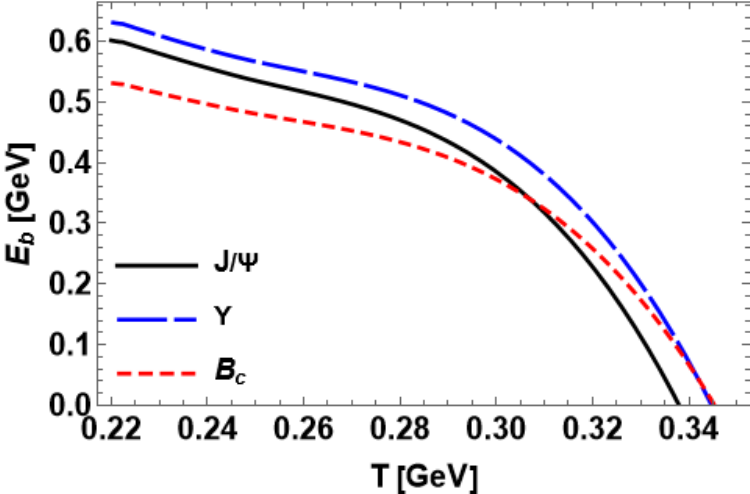}
    \caption{Plot of the dissociation energy defined in Eq.~(\ref{Eb}) for the $J/\psi$, $\Upsilon$ and $B_c $ mesons as a function of the temperature $T$ .}   
    \label{fig1}
\end{figure}

Again, for completeness, we start by presenting in Fig.~\ref{fig1} the behavior of the dissociation energy in Eq.~(\ref{Eb}) of the $J/\psi, \Upsilon$ and $B_c $ mesons as a function of the temperature. As can be seen, $E_b$ decreases as the temperature increases. Interestingly, these vector quarkonia have close dissociation temperatures, but with the $J/\psi $ dissociating at a slightly smaller temperature. 

\begin{figure}[htbp!]
    \centering
    \includegraphics[width=0.6\textwidth]{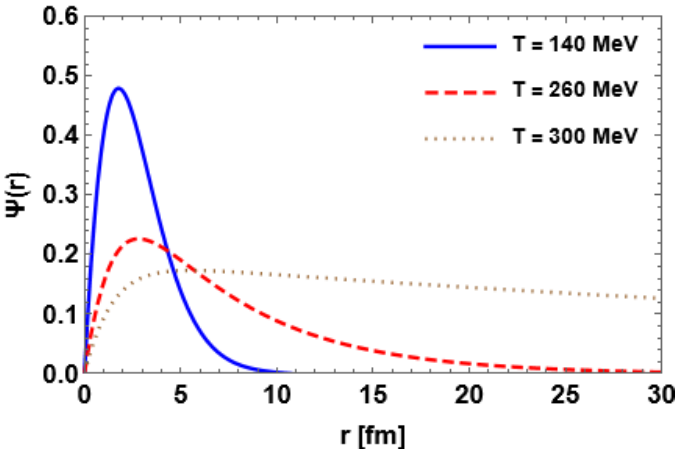}
    \caption{Wave function of the bound state solution associated to  the $J/\psi $ as a function of the radial coordinate for different values of temperature.}   
    \label{figwf}
\end{figure}

Besides, in order to illustrate how the eingenstates behave with the temperature, in Fig.~\ref{figwf} the wave function of the bound state solution associated to the $J/\psi $ for different values of temperature is plotted. We can interpret it as follows. At lower temperatures, the wave function has a more prominent peak at lower values of the radial coordinate. 
Higher temperatures engender the spread of the peak along a bigger range of $r$, with its maximum value being shifted to larger values of $r$. 
At the same time, in the QGP phase quarks and gluons become deconfined color charges, leading to a color screening, which is characterized by the color screening radius $r_D$ determining the range of the strong interaction~\cite{Satz:2006uh}. Noting that $r_D$ is inversely proportional to the density of charges and decreases with increasing temperature, as a result the $Q \bar Q$ interaction becomes even more short-ranged.
Then, when the system approaches the dissociation temperature, the peak of the wave function becomes negligible, the system reaches a distance enough to break the string and form two light-heavy mesons, and the bound state no longer exists. 

We summarize in Table~\ref{tab:quarkonia_predictions_temp} the dissociation temperature for quarkonium states in different sectors previously considered. The dissociation temperature is also given in units of the critical temperature at which the Quark-Gluon Plasma (QGP) is formed, $T_c$, which here is considered $T_c=170 $ MeV. The dissociation of these quarkonia happens within the range $(1.91 - 2.06) \ T_c$. However, it should be stressed that the absolute values of $T_{dis}$ and $T_{dis}/T_c$ depend on the choice of the parameterization for the potential, the Debye mass, parameters set, and $T_c$. In particular, for the case of $J/\psi(1S)$ our result is in accordance with that obtained in Refs.~\cite{Song:2011xi,Gauss}. In general we have a slight decrease of $T_{dis}$ for radially excited states, and a mild augmentation for orbitally-excited quarkonia.

\begin{table}[htbp!]
\centering
\caption{Summary of theoretical predictions of the dissociation temperature  $T_{dis}$ for  mesons in the charmonium, bottomonium and bottom-charmonium sectors of the spectrum. In the last column the dissociation temperature is given in units of the critical temperature $T_c = 170$ MeV. } 
\label{tab:quarkonia_predictions_temp}
\begin{tabular}{c c c c }
\toprule
\hline
$(Q \bar Q)$ & $N^{2S+1}L_J$ & $T_{dis}$ [MeV] &  $T_{dis}/T_c$ \\
\midrule
\hline
$\eta_c(1S)$   & $1^1S_0$ & 342 & 2.01 \\
$J/\psi(1S)$   & $1^3S_1$ & 338 & 2.00 \\
$\chi_{c0}(1P)$ & $1^3P_0$ & 345.6 & 2.03 \\
$\chi_{c1}(1P)$ & $1^3P_1$ & 348.2 & 2.05\\
$h_c(1P)$      & $1^1P_1$ & 348.4 & 2.05 \\
$\chi_{c2}(1P)$ & $1^3P_2$ & 350.1 & 2.06 \\
$\eta_c(2S)$   & $2^1S_0$ & 328.3 & 1.97 \\
$\psi(2S)$   & $2^3S_1$ & 325& 1.91 \\
$\chi_{c0}(2P)$ & $2^3P_0$ & 345 & 2.03 \\
$\chi_{c1}(2P)$ & $2^3P_1$ & 345.9 & 2.03\\
$h_c(2P)$      & $2^1P_1$ & 348 & 2.05 \\
$\chi_{c2}(2P)$ & $2^3P_2$ & 348.2 & 2.05\\
$\eta_b(1S)$ & $1^1S_0$ & 345 & 2.03 \\
$\Upsilon(1S)$ & $1^3S_1$ & 344 & 2.03 \\
$\chi_{b0}(1P)$   & $1^3P_0$ & 337 & 1.98 \\
$\chi_{b1}(1P)$   & $1^3P_1$ & 335 & 1.97 \\
$h_b(1P)$       & $1^1P_1$ & 348 & 2.04 \\
$\chi_{b2}(1P)$   & $1^3P_2$ & 336 & 1.98 \\
$\Upsilon(2S)$ & $2^3S_1$ & 329 & 1.96 \\
$\chi_{b0}(2P)$   & $2^3P_0$ & 333.9 & 1.97 \\
$\chi_{b1}(2P)$   & $2^3P_1$ & 330.6 & 1.94 \\
$h_b(2P)$       & $2^1P_1$ & 345 & 2.03 \\
$\chi_{b2}(2P)$   & $2^3P_2$ & 329.7 & 1.94 \\
$B_{c}$(1S) &  $1^1S_0$ & 347.2 & 2.04 \\
$B_{c}$ (1S)& $1^3S_1$ & 345.6 & 2.03 \\
$B_{c}$(2S) &  $2^1S_0$ & 326 & 1.92 \\
$B_{c}$ (2S)& $2^3S_1$ & 328 & 1.93 \\
\hline
\bottomrule
\end{tabular}
 \end{table}

\subsubsection{Tetraquark states}

\begin{figure}[htbp!]
    \centering
    \includegraphics[width=0.6\textwidth]{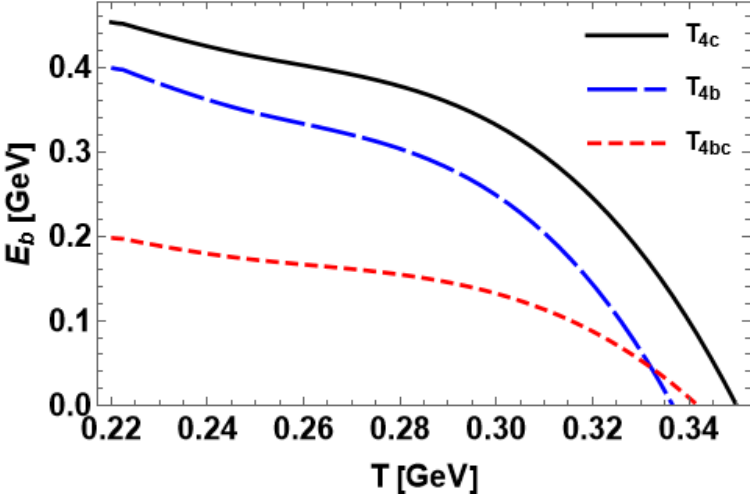}
    \caption{Plot of the dissociation energy defined in Eq.~(\ref{Eb}) for the ground tetraquark states $1^1S_0$ in different sectors denoted as $T_{4c}, T_{4b}$ and $T_{2b2c}$ as a function of the temperature $T$ .}   
    \label{fig1t}
\end{figure}

Now we analyze the tetraquarks properties in the medium. In Fig.~\ref{fig1t} is plotted the behavior of the dissociation energy in Eq.~(\ref{Eb}) for the ground tetraquark states $1^1S_0$ denoted as $T_{4c}, T_{4b}$ and $T_{2b2c}$, as a function of temperature. Like in the case of quarkonia, $E_b$ decreases as the temperature increases. But curiously noting that $E_b$ is smaller and its reduction is slower, then these states experience dissociation at similar temperatures with respect to the quarkonia. In the present case, the $T_{4c} $ dissociates at a slight higher temperature, and the $T_{4b}$ presents a vanishing dissociation energy at the smallest $T_{dis}$. 

The results for the predicted dissociation temperatures of the tetraquarks in different sectors and with different quantum numbers are provided in Table \ref{tab:quarkonia_predictions_temp}. The dissociation of these states happens within the range $(1.95 - 2.06) \ T_c$, similarly to the one found for the quarkonia. We see that the effect of a smaller $T$-dependent interacting potential between the diquarks and anti--diquarks gives a smaller dissociation energy but with a slower decreasing with temperature.

\begin{table}[htbp!]
\centering
\caption{Summary of theoretical predictions of the dissociation temperature $T_{dis}$ 
for the tetraquarks in different sectors and with different quantum numbers. In the last column the dissociation temperature is given in units of the critical temperature $T_c = 170$ MeV. }
\label{tab:tetraquark_predictions_temp}
\begin{tabular}{c c c c }
\toprule
\hline
$T_{4Q}$  & $N^{2S+1}L_J$ &$T_{dis}$ [MeV]  &  $T_{dis}/T_c$ \\
\midrule
\hline
$T_{4c}$ & $1^1S_0$ & 349.5 & 2.06 \\
$T_{4c}$ & $1^3S_1$ & 349 & 2.06 \\
$T_{4c}$ & $1^3P_0$ & 334.5 & 1.97 \\
$T_{4c}$ & $1^3P_1$ & 332.1 & 1.95\\
$T_{4c}$ & $1^1P_1$ & 334.0 & 1.96\\
$T_{4c}$ & $1^3P_2$ & 333.2 & 1.96\\
$T_{4b}$ & $1^1S_0$ & 338.5 & 1.99\\
$T_{4b}$ & $1^3S_1$ & 336.6 & 1.98\\
$T_{4b}$ & $1^3P_0$ & 344.7 & 2.03\\
$T_{4b}$ & $1^3P_1$ & 345.1 & 2.03 \\
$T_{4b}$ & $1^1P_1$ & 344 & 2.03 \\
$T_{4b}$ & $1^3P_2$ & 345.1 &2.03 \\
$T_{2b2c}$ & $1^1S_0$ & 339.3 & 2.00\\
$T_{2b2c}$ & $1^3S_1$ & 341.8 & 2.01\\
$T_{2b2c}$ & $1^3P_0$ & 341.5 & 2.00\\
$T_{2b2c}$ & $1^3P_1$ & 345 & 1.99\\
$T_{2b2c}$ & $1^1P_1$ & 343 & 2.00 \\
$T_{2b2c}$ & $1^3P_2$ & 342 & 2.01 \\
$T_{4c}$ & $2^1S_0$ & 331 & 1.95 \\
$T_{4c}$ & $2^3S_1$ & 326.9 & 1.92 \\
$T_{4c}$ & $2^3P_0$ & 331 &  1.95 \\
$T_{4c}$ & $2^3P_1$ & 330.1 & 1.94 \\
$T_{4c}$ & $2^1P_1$ & 333 & 1.96\\
$T_{4c}$ & $2^3P_2$ & 330.9 & 1.95\\
$T_{4b}$ & $2^1S_0$ & 336 & 2.01\\
$T_{4b}$ & $2^3S_1$ & 338 & 1.99\\
$T_{4b}$ & $2^3P_0$ & 330 & 1.94\\
$T_{4b}$ & $2^3P_1$ & 328.9 & 1.93 \\
$T_{4b}$ & $2^1P_1$ & 331 & 1.95 \\
$T_{4b}$ & $2^3P_2$ & 329 & 1.94\\
$T_{2b2c}$ & $2^1S_0$ & 335 & 1.97\\
$T_{2b2c}$ & $2^3S_1$ & 336.9 & 1.98\\
$T_{2b2c}$ & $2^3P_0$ & 338 & 1.99\\
$T_{2b2c}$ & $2^3P_1$ & 328 & 1.93\\
$T_{2b2c}$ & $2^1P_1$ & 329.5 & 1.94\\
$T_{2b2c}$ & $2^3P_2$ & 336.9 & 1.98\\
$T_{4c}$ & $3^1S_0$ & 329 &  1.94\\
$T_{4c}$ & $4^1S_0$ & 328 &  1.93\\
\hline
\bottomrule
\end{tabular}
 \end{table}

In the end, these findings provide us the following characterization: when the fireball has a temperature which is higher than about twice the critical temperature, tetraquark and quarkonia states cannot be formed in the thermal medium due to the Debye screening mechanism. Notwithstanding, when the system reaches lower temperatures than $2T_c$ all these states can be formed.

\subsubsection{Comparison with other models} 

Before ending this Section, it is worthwhile to analyze and compare our approach with other models that incorporate in the finite-temperature potential not only the real part of the in-medium interaction, but also an imaginary one. 
As highlighted in some studies such as Ref.~\cite{Laine:2006ns} (see also~\cite{Song:2020kka,Armesto:2024zad,Gauss}), a Debye-screened potential describing the binding of a quark-antiquark pair at elevated temperatures can be derived by defining a gauge-invariant Green's function, which can be calculated to the first non-trivial order in Hard Thermal Loop resummed perturbation theory. This potential naturally develops an imaginary component, indicating that thermal effects may yield a finite width for the quarkonium, making the quarkonium peak in production rates wider than at zero temperature. 
Physically, this phenomenon stems from the energy transfer between low-frequency ($q_0$) gauge fields, which facilitate the static interaction, and the "hard" particles in the thermal plasma, possessing momenta ($\mathbf{q}$) comparable to the temperature. This phenomenon is also known as Landau damping. 
On a technical level, it arises from the cut contribution to the gluon spectral function when $|q_0| < |\mathbf{q}|$. 
Focusing on the qualitative behavior of a bottomonium system and treating the imaginary component perturbatively,  Ref.~\cite{Laine:2006ns} has found that as the inverse of the Debye mass (i.e. the Debye radius) becomes significant, the imaginary part of the potential acquires non-vanishing values and the wave function spreads out. As a consequence, the binding energy rapidly diminishes and at the same time the width increases. At a some point, the width exceeds the binding energy and the bottomonium ``melts". 
Using a similar potential, Ref.~\cite{Song:2020kka} studied  the~$J/\psi$ system, but taking two constraints: first by solving the Schr\"odinger equation and second through the QCD sum rule approach. 
The findings suggest that the binding energy is minimally affected by the strength of the imaginary potential, whereas the width varies proportionally with the strength of the imaginary potential. Besides, the $J/\psi$ dissociation occurs when binding energy and thermal width become comparable in magnitude. Beyond this temperature, the state loses its identity; and the sum rule analysis becomes unstable.

Thus, benefiting from the analysis above,  one can infer that the inclusion of an imaginary part in the potential of our model would produce the following effects on our results. Assuming that the imaginary potential does not affect the real part of the solutions of the Schr\"odinger equation, the binding energies of the systems studied here would remain practically the same, as in the plots depicted in Figs.~\ref{fig1} and~\ref{fig1t}. The main difference would be in the fact that the states would acquire a thermal width, influencing on the shape of the peak associated to the state in production rates observed by the experiments. Also, 
the dissociation temperatures shown in Tables~\ref{tab:quarkonia_predictions_temp} and~\ref{tab:tetraquark_predictions_temp} would probably be affected if the criterium of determination of the dissociation temperature would be adopted as in Refs.~\cite{Laine:2006ns,Song:2020kka}, namely: the temperature at which the binding energy and thermal width have comparable magnitudes. In that regard, the values of $T_{dis}$ might diminish with respect to the ones reported in the mentioned tables. Nevertheless, since the strength of the imaginary potential depends on the model and the constraints employed, it is not a trivial task to have a precise prediction of this effect. We postpone it for a future work. Notwithstanding,  emphasizing  our interest on a qualitative description of the thermal behavior of the $T_{4Q}$'s, the conclusions discussed previously would remain qualitatively valid even with the presence of an imaginary potential.

\section{Discussion and concluding remarks}

The main aim of this study has been the analysis of the properties of exotic states and quarkonia in heavy-ion collisions, in particular by characterizing how they behave in the thermal medium. We have used a quark model that can reasonably describe the meson spectrum in vacuum by solving the Schrodinger equation for the Cornell potential. We have fitted the set of parameters employing the $\chi^2$ (chi-square) minimization method, using as input the experimental masses of the ground states and their excitations. This formalism has been expanded to construct tetraquarks states, by understanding how the color structure would change in order to couple a $1^3 S_1$ axial $QQ$ pair to a $1^3 S_1$ axial $\bar{Q}\bar{Q}$ pair. 

For our predictions of the spectrum of the tetraquark states in the vacuum, we have found that the structures $X(6600)$ (or $X(6552)$), $X(6900)$ and $X(7200)$ reported  by the CMS Collaboration are compatible with our predictions for the radially-excited $T_{4c}(n^1S_0)$ configurations with $n=2,3,4$. However, one should notice that there is not yet a definite theoretical description for the experimental data. For example, Ref.~\cite{Belov:2024qyi} proposed the interpretation of $X(6900)$ as a $T_{4c}\ [2^{++}(2S)]$. On the other hand, Ref.~\cite{Zhu:2024swp} interpreted the triplet masses by a typical Regge trajectory of radial excitations, where the $X(6900)$ is a $n=3$ radially excited state, as in our case. 
In the end, there are several exploratory pictures which are by no means conclusive. 
Therefore, to get more refined and accurate results, improvements in our framework are needed, as for instance: more realistic interaction potentials, exploration of orbitally and radially excited diquark configuration, and so on. 

On experimental grounds, ongoing and future data samples will help us to elucidate questions, such as intrinsic configuration, spin-parity, interference and possible new states. And HICs appear as a promising scenario, where the comprehension of how the exotic states behave in the thermal medium plays a relevant role. 
On this subject, we have introduced the temperature dependence through a temperature-dependent potential, which allowed us to predict the dissociation temperatures for mesons and tetraquarks. Our findings suggest that, as a consequence of the Debye screening mechanism, they cannot be formed in the thermal medium when the system has a temperature higher than about twice the critical temperature. However, at sufficiently smaller temperatures these states will be formed in a similar range of $T$. 

We emphasize that the quantitative behavior of the dissociation energy and the values of $T_{dis}$ and $T_{dis}/T_c$ are strongly dependent on the choice of the parametrization. Notwithstanding, the present analysis can provide some insights into the structure of the recently observed tetraquark states, their dissociation mechanism states caused by a thermal medium, and the temperature range at which the tetraquark states might be formed. These findings may be useful to better understand the exotics in HICs.

Finally, we deserve a comment on the model’s scope and limitations, and whether our framework could be applied to other types of exotics. It is well known that potential models have been successfully used to describe quarkonia, and more recently exotic hadronic states. As noted in~\cite{Armesto:2024zad}, heavy tetraquarks can be interpreted as electromagnetic molecule-like systems, provided they satisfy two key conditions for the Born-Oppenheimer approximation to apply, namely: (i) the non-relativistic approximation must hold, meaning that the velocity of the heavy quarks around the meson’s center-of-mass should be small; (ii) the binding energy must be significantly smaller than the $\Lambda_{QCD}$. Accordingly, the heavy quarks act as the slow degrees of freedom, while light quarks and gluons play the role of fast degrees of freedom, which are then encapsulated into an effective potential, derived in the limit where the heavy quarks are treated as static, infinitely massive sources. That said, the present approach can be naturally expanded to evaluate tetraquark systems with open heavy flavor, like $c c -\bar b \bar c $ and $ b b - \bar b \bar c$, since they comply with the two constraints mentioned above. Other cases involving light-quark content should be considered carefully within our formalism. However, it is worthy noticing that Ref.~\cite{Armesto:2024zad} applied the diquark-antidiquark model  $c q - \bar c \bar q$ for the $X(3872)$ state, taking into account for the small separation distances, where the interaction is principally governed by one-gluon exchange, the simplification of focusing exclusively on the single octet color channel. A similar choice has also been done in other studies; see for example~\cite{Lundhammar:2020xvw} and references therein. In this sense, we plan to explore other loosely bound or mixed-flavor tetraquark candidates, as well as other distinct assumptions to the compact axial diquarks, in order to have a more thorough understanding of this matter and to identify where the model succeeds and where it falls short.

\section{Acknowledgements}

We would like to thank Franz F. Sch\"oberl for providing us
with his Mathematica notebook, which has been taken as basis for the code used for solving the Schrödinger equation. 
This work was partly supported by the Brazilian Brazilian CNPq (L.M.A.: Grants No. 400215/2022-5, 308299/2023-0, 402942/2024-8), CNPq/FAPERJ under the Project INCT-F\'{\i}sica Nuclear e Aplica\c c\~oes (Contract No. 464898/2014-5) and CAPES (V.S.S.).


\end{document}